COMPARATIVE STUDY OF PHASE TRANSITIONS IN POLYCRYSTALLINE AND EPITAXIAL BaTiO$_3$ THIN FILMS BY MEANS OF SPECIFIC HEAT MEASUREMENTS

B.A.Strukov[1], S.T.Davitadze[1], V.V.Lemanov[2], S.G.Shulman[2], Y.Uesu[3] and S.Asanuma[3]

[1]*Lomonosov Moscow State University, Moscow 119992, Russia*
 *e-mail: bstrukov@mail.ru*
[2]*Ioffe Physico-Technical Institution RAS, S.-Petersburg 194021, Russia*
[3]*Waseda University, Tokyo 169-8555, Japan*

Abstract   The experimental data on the thermal properties of the nanostructured perovskite ferroelectrics are presented and analysed. The ability of the modified 3ω method for specific heat measurements is discussed and illustrated by the thermal properties of the thin polycrystalline (BaTiO$_3$ on fused SiO$_2$, 50-1100 nm) and epitaxial (BaTiO$_3$ on MgO, 50-500 nm) films. The sharp decrease of the transition temperature, surplus entropy and spontaneous polarization were obtained for the polycrystalline films. The critical thickness and crystallite size are estimated for this case from the variation of both crystallite sizes at constant thickness and film thickness at constant crystallite size. The similarity of specific heat, phase transition excess entropy evolution in polycrystalline films and nanograined ceramics (the latter was obtained in [12] ) is revealed. The data on the specific heat of epitaxial films show the extremely diffused anomaly near 400K, the phase transition has the weak tendency of shifting to higher temperatures with the decreasing of the film thickness. No anomalies were detected for the thinnest (50 nm) film up to 480K.

   In the field of ferroelectricity the problem of finite-size effects received the powerful practical interest because of the number of applications of the ferroelectric nanostructured devices [1-3]. It is very important to know the limit of presence of ferroelectric properties and phase transition when the size of devices is reduced. Thin films are the most interesting objects in this respect;  as well for applications  it is  important to know the dependence of physical properties of ferroelectric films and ceramics on the crystallite size.  Therefore the finite-size effects in ferroelectrics have both the practical importance and fundamental scientific sounding and several recent review papers and books on the subject and its development throw much light on the problem [4-5]. It should be noted that these sources are accented mainly on consideration of ferroelectric oxides films as elements of memory devices and therefore the polarization switching processes, coercive electric field and related phenomena are considered there as subjects of the main interest. In this paper we shall concentrate our attention mainly on the thermal properties of the nanostructured perovskites directly related to ferroelectric phase transitions. Perovskite ferroelectric oxides like BaTiO$_3$, PbTiO$_3$, SrTiO$_3$, (Ba,Sr)TiO$_3$ etc are the most attractive for both applications and as models for finite-size  effect study because of their relatively simple crystal structure and well developed methods of  manufacturing.

Although the switching parameters and dielectric properties of the ferroelectric nanostructures have a prevalence importance at the moment, the critical phenomena and phase transition singularities seem to have a fundamental sounding for the subsequent understanding of the size-effects nature in these objects. It should be noted that the experimental data obtained by means of dielectric measurements in films might be essentially distorted by the depletion layers with a low dielectric constant, arising near electrodes [6,7]. Since we need an information about *volume* properties of the nanoparticles or films, the "non-electrical" data can be useful for the characterization of the phase transitions in these ferroelectric nanosized objects. We consider specific heat measurements as an important method to obtain the direct information about transition temperatures and phase transition behavior; moreover it gives a chance to get indirectly a spontaneous polarization temperature dependence in the vicinity of the transition point.

The temperature dependence of the specific heat of free-standing (without substrate) ferroelectric films was analysed in [8-10] within Landau-Ginzburg-Devonshire (LGD) theory phenomenological approach including in the free energy density of the term proportional to $|gradP|^2$ together with a boundary condition involving an extrapolation length $\delta$. The finite discontinuity in the specific heat was found at the film critical temperature $T_c$. The jump of the film specific heat, excess entropy and $T_c$ position were found to decrease together with the decreasing of a film thickness.

Methodically there are no special problems with the application of the scanning and adiabatic calorimetry for investigation of the size effects in powder (usually fine particles compacted into pellets) and nanograined ceramics. The rather sharp decrease of the Curie temperature as well as of excess energy of the phase transition were revealed for $BaTiO_3$ (BTO) and $PbTiO_3$ (PTO) fine particles [11] and BTO ceramics [12-14] for the diminishing size of particles and crystallites. These effects were considered as an evident manifestation of the theoretically predicted size effect. The obvious smearing of the phase transition was observed in the both cases; for the ceramics the problems of correlation of dielectric and thermal properties was discussed [12].

The first measurement of the specific heat temperature dependence for the epitaxial BTO films on $SrTiO_3$ substrate was undertaken by A.Onodera et al. with usage of the *ac*-method [15]; the data revealed the small and diffused anomalies for the 6 and 200 nm thick monocrystalline films. The thermal hysteresis - a visible difference of the specific heat values on heating and on cooling - was detected. Authors supposed the significant increasing of the transition temperature in comparison with the bulk samples which was attributed to influence of the lattice misfit strain between films and substrate.

The next attempts to measure specific heat of the thin ferroelectric films sputtered on the substrate were undertaken when the 3ω dynamical method was applied to the analysis of thermal properties of the composite samples dielectric film – massive substrate [16,17]. It was shown that if the thermal contrast K between film and substrate ( $K=c_f\lambda_f/c_s\lambda_s$, $c$ is specific heat of the film (f) and substrate (s) ) is small enough, the direct measurement of the film specific heat is possible down to 20 nm film thickness [16]. Therefore $BaTiO_3$ films on the fused $SiO_2$ and monocrystalline MgO as substrates seem to be convenient match to measure specific heat of the polycrystalline and epitaxial films (K=0.1 for the $BaTiO_3$ / $SiO_2$ and 0.2 for $BaTiO_3$ / MgO couples ).

$BaTiO_3$ polycrystalline films were grown on the fused quartz by rf magnetron sputtering at the different substrate temperatures in atmosphere $Ar/O_2 \approx 0.8/0.2$. The rate of sputtering was about 300 nm/h. AFM analysis of the film surface showed that the size of monocrystalline regions depends on the temperature of substrate in time of sputtering. For the temperature of substrate near $800^0C$ it is to be about 110-150 nm in the interval of film thickness 100-1100 nm and decreases sharply with the further decreasing of a thickness. With decreasing of the substrate temperature the average size of the monocrystalline regions was reduced, so we had a chance to modify both the film thickness and crystalline size separately. The columnar-like structure was found for all these films. The epitaxial films were grown on MgO single crystal substrate ($a_s$ = 4.213 A) by the pulsed laser deposition method. The surfaces of all films were controlled by AFM. In polycrystalline films the mosaic structure was evident; but in the case of epitaxial films no grains of mosaic were observed revealing the perfect "epitaxial coupling" between film and substrate. AFM images of 100 nm BTO films deposited on fused $SiO_2$ and MgO are shown at Fig.1.

The temperature dependence of the specific heat of polycrystalline $BaTiO_3$ films are presented at Fig. 2a (variation of the thickness for an approximately constant column diameter) and at Fig. 2b (variation of the column diameter for a constant film thickness). In the both cases the sharp decrease of the transition temperature and excess entropy of the phase transition is evident (Fig. 3a, 3b); the phase transition becomes diffused when the film thickness declined. The temperature dependence of the spontaneous polarization can be taken out from Fig.3 by the well known LGD theory relations and is presented at Fig. 4a, 4b. It is clearly seen that spontaneous polarization becomes smaller for the thinner polycrystalline films; it is important that in these experiments we obtain (although indirectly) the real volume polarization whereas in the direct measurements of $P_s$ from hysteresis loop or switching current one can not control the nonswitching regions of the samples. It is known that LGD theory predicts that the shift of the transition temperature is $T_b - T_c = Ad^{-1}$ where $T_b$ is the Curie temperature for a bulk sample, $d$ is the crystal thickness or grain size and $A$ – combination of LGD free energy coefficients. So we can use the obtained data for evaluation of critical thickness and critical grain size from the condition $T_c = 0$. From this condition we have for films $d_{crit} \approx 2.5$ nm and for grains $d_{crit} \approx 8$ nm. It should be noted that in general the evolution of the phase transition in polycrystalline thin films, as considered from the thermal properties, looks rather similar to that observed in nanograined ceramics [12]. Evidently in both cases the observed changes of the transition temperature, excess entropy and spontaneous polarization can be attributed to an intrinsic size effect. It was shown in [12] that the residual stresses resulting from the tetragonal deformation play a minor role on the behavior of the fine grained ceramics; the critical size of the grains in nanocrystalline ceramics was estimated as 10-30 nm. In our case of the polycrystalline thin films on the fused quartz, the additional two-dimensional stresses arising from an interaction between film and substrate seem also to be almost completely relieved. The observed smearing of phase transition in the fine particles, nanoceramics and polycrystalline thin films can be described by taking into account the distribution of the grain sizes which exists in all nanomaterials and results in the corresponding distribution of transition temperatures [18].

It was revealed that the quite another type of thermal behavior is typical for the epitaxial films $BaTiO_3/MgO$. The temperature dependence of the specific heat and spontaneous polarization of films with the different thickness is presented at Fig. 5,6. It is seen that there is a weak tendency of increasing of the transition temperature which can be identified as a maximum of the heavily diffused specific heat anomalous part (see insert at Fig.5). It is also remarkable that the film with the thickness 50 nm shows no anomalous change up to 480K (that is an upper temperature limit for our method at present). We consider this result as an evident competition of the two effects, one is the size-effect theoretically predicted for free-standing films (declining of $T_c$ with decreasing of the film thickness) and the second is a mismatch between the MgO lattice parameter $a_s$ and the cubic cell constant $a_o$ of the free-standing $BaTiO_3$ film, clamping the film in two directions. In fact the lattice mismatch is very large for $BaTiO_3/MgO$ couple: the misfit strain defined as $u_m = (a_s - a_o)/a_s + (\alpha_s -$ $\alpha_o)(T - T_g)$ (here $\alpha_s$ and $\alpha_o$ are the thermal expansion coefficients of the substrate and film, $T_g$ is the crystal growth temperature) is about 5%.

It was shown at [19] that when the substrate lattice parameter is larger than the film lattice parameter and the film thickness is larger than the critical value $h_c \approx$ 30-50 nm, the film may partially relax at the growth temperature by the generation of misfit dislocations. It leads to a value of the "effective" substrate lattice parameter $a_{s\ eff} < a_s$ so that the BaTiO$_3$ film on MgO substrate has in fact a tensile strain of near 0.57% [20]; if one takes into account the temperature-misfit strain phase diagram for epitaxial (001) BaTiO$_3$ film revealed in [21], the phase transition is supposed to be at about 600K. Therefore we are forced to assume that the further release of the tensile mechanical stresses takes place at least for the films thicker than 50 nm. The Table sums up the data, obtained for polycrystalline (with completely relieved stresses) and epitaxial films; here $T_c$ is the transition temperature of the epitaxial films, $\Delta T_c$ (the second column) – displacement of $T_c$ due to the size effect (in polycrystalline films), $\Delta T_c$ (the third column) - displacement of $T_c$ due to both size effect and mismatch effect for the each thickness of epitaxial films. The last column gives the total residual strain resulting in the observed shift of $T_c$. The relationship between $\Delta T_c$ and $u_m$ is taken from [21].

In summary we studied the thermal properties of BaTiO$_3$ polycrystalline and epitaxial films by means of the modified 3ω method. The clear size effect, similar to one obtained in the nanostructured ceramics, was detected and the critical film thickness and grain size were estimated in polycrystalline BaTiO$_3$/SiO$_2$ films. Through the temperature dependence of the specific heat of epitaxial films BaTiO$_3$/MgO we revealed the misfit strains which were found to decrease when the film thickness grows up.

This work was partly supported by RFBR grants (projects 05-02-16873 and 06-02-16664)

REFERENCES

1. J.F.Scott, Ferroelectric Review **1**, 1-129 (1998).
2. J.F.Scott, Ferroelectric Memories (Springer-Verlag, Berlin, 2000).
3. M.Dawber, K.M.Rabe and J.F.Scott, arxiv: cond-mat/0503372 v1, 15 Mar.2005.
4. Eds C.P.de Araujo, J.Scott and G.W.Taylor, Ferroelectric Thin Films: Synthesis and Basic Properties (Gordon and Breach Publishers, Amsterdam, 1996).
5. Eds H.Ishiwara, M. Okuyama and Y.Arimoto, Ferroelectric Random Access Memories (Springer-Verlag, Berlin, 2004).
6. O.G.Vendik and S.P.Zubko, J.Appl.Phys. 88, 5343 (2000).
7. A.Lookman, R.M.Bowman and J.M.Gregg, J.Appl.Phys. 96, 555 (2004).
8. C.Wang and S.Smith, J.Phys.: Cond.Matter 7, 7163 (1995).
9. S.V.Pavlov and O.Yu.Poliakova, Vestnik Mosk. Univ., Ser. Phys. Astronom. №1, 44 (2000).
10. L.-H.Ong, J.Ostman and D.Tilley, Phys.Rev. B, 63, 144109 (2001).
11. W.Zhong et al., J.Phys.: Cond.Matter 5, 2619 (1993).
12. Z.Zhao et al., Phys.Rev. B 70, 024107 (2004).
13. M.H.Frey and D.A.Payne, Phys.Rev. B 54, 3158 (1996).
14. V.Tura et al., Jpn.J.Appl.Phys. 37, 1950 (1998).
15. A.Onodera et.al., J.Europ.Ceramic Soc. 19, 1477 (1999).
16. S.T.Davitadze et al., Solid State Phys., 42, 111 (2000).
17. B.A.Strukov et.al., J.Phys.: Cond.Matter 15, 4331 (2003).
18. M.D.Glinchuk and P.I.Bykov, J.Phys.: Cond.Matter., 16, 6779 (2004).
19. J.S.Speck and W.Pompe, J.Appl.Phys. 76, 466 (1994).
20. F.He and B.O.Wells, arxiv: cond-mat/0511518, 21 Nov. 2005.
21. N.A.Pertsev et al., Phys.Rev.Lett. 80, 1988 (1998).

Captions to figures to the paper of B.A.Strukov et al.

Fig.1. AFM image of BTO films deposited on fused $SiO_2$ (a) and MgO (b).

Fig.2. Temperature dependence of the specific heat for polycrystalline $BaTiO_3/SiO_2$ films with different thickness (a) and different grain size (b).

Fig.3. Temperature and excess entropy of the ferroelectric phase transition in polycrystalline $BaTiO_3/SiO_2$ films with different thickness (a) and different grain size (b).

Fig.4. Temperature dependence of spontaneous polarization in polycrystalline $BaTiO_3/SiO_2$ films with different thickness (a) and different grain size (b).

Fig.5. Temperature dependence of the specific heat of $BaTiO_3/MgO$ epitaxial films. Inset – $T_c$ and excess entropy *vs* film thickness.

Fig.6. Temperature dependence of spontaneous polarization for $BaTiO_3/MgO$ epitaxial films.

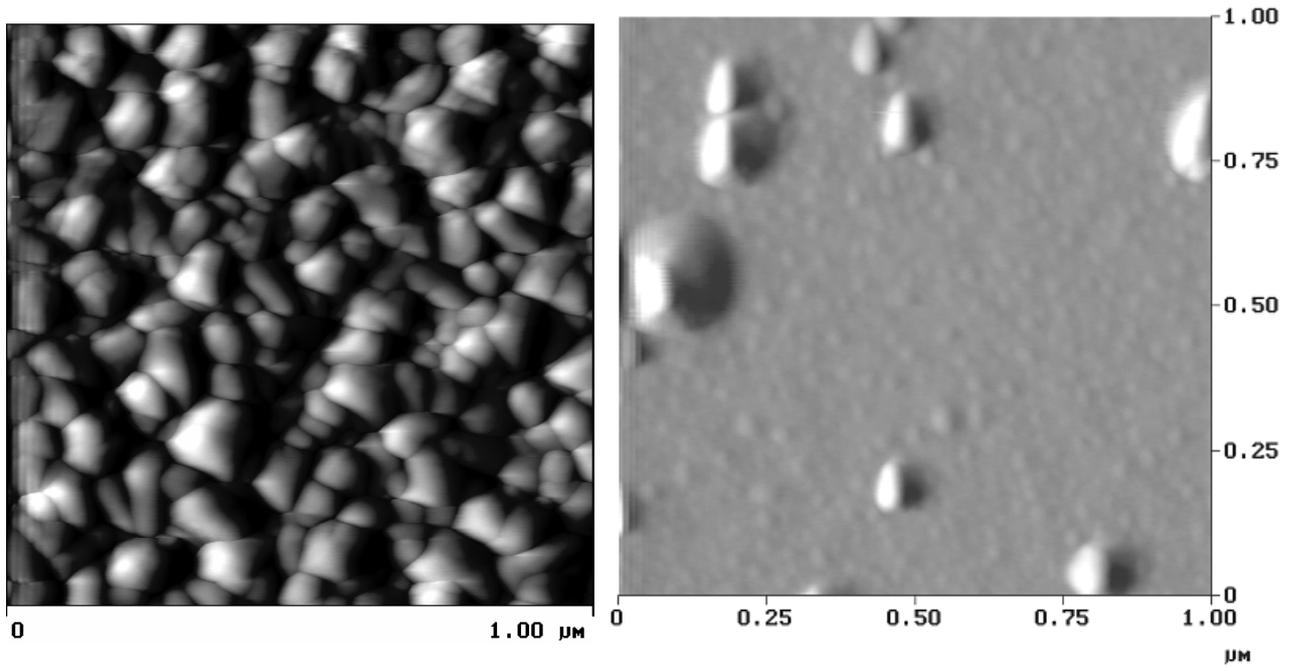

Fig.1          (a)                              (b)

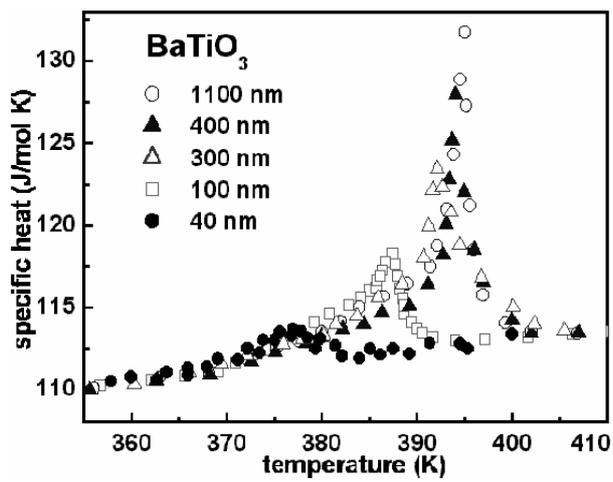
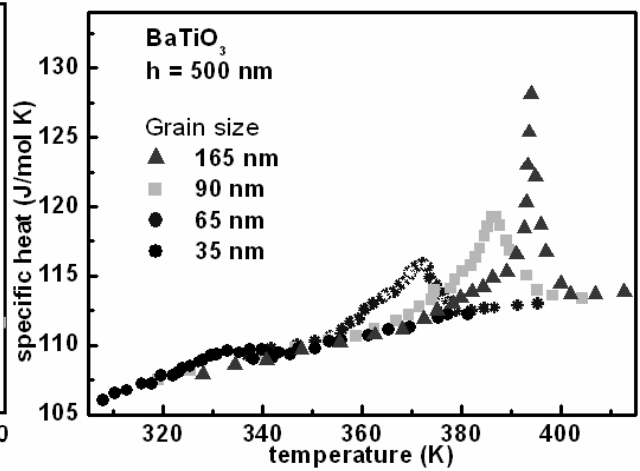

Fig 2          (a)                              (b)

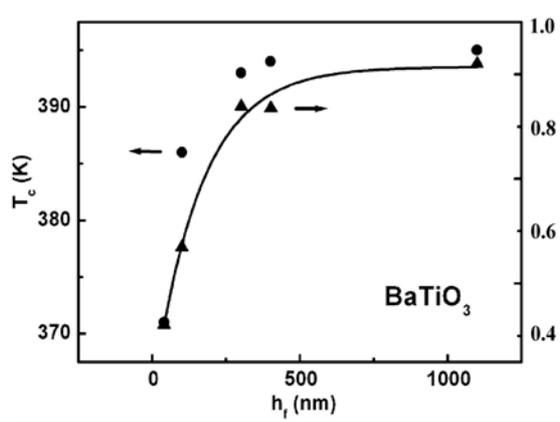
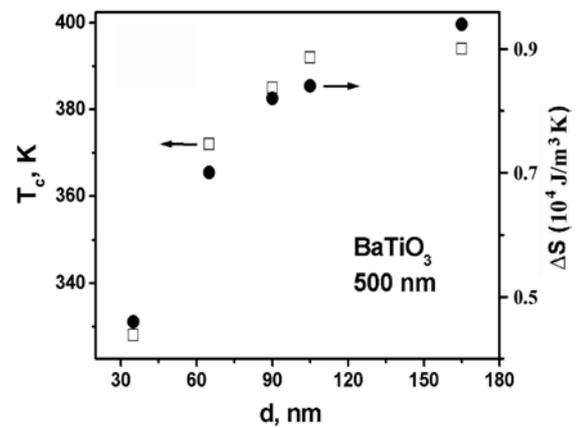

Fig. 3          (a)                              (b)

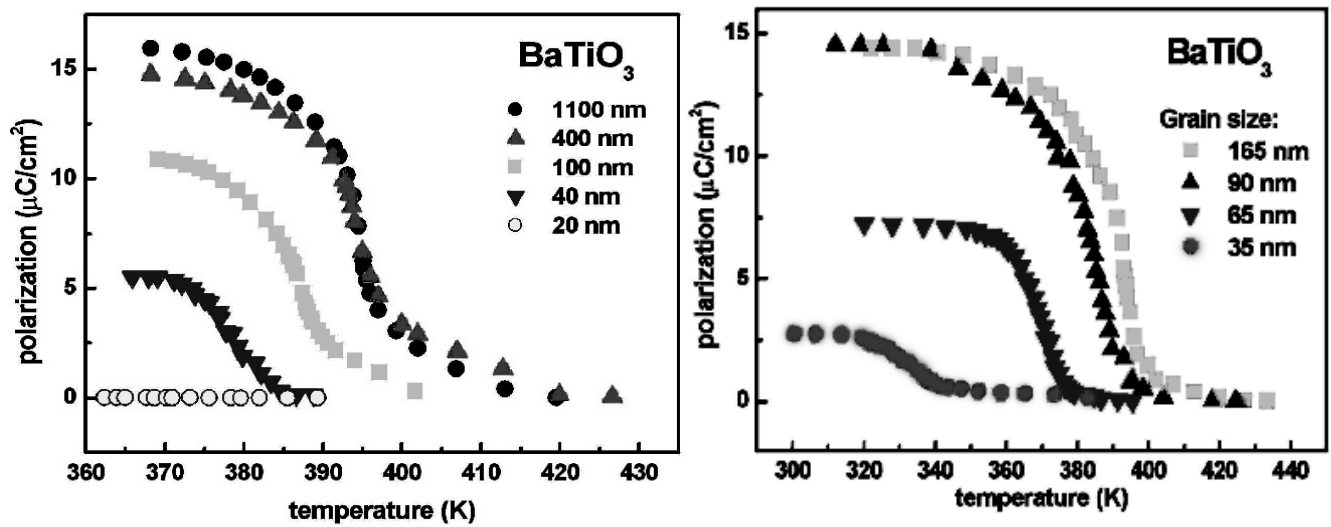

Fig 4.           (a)                              (b)

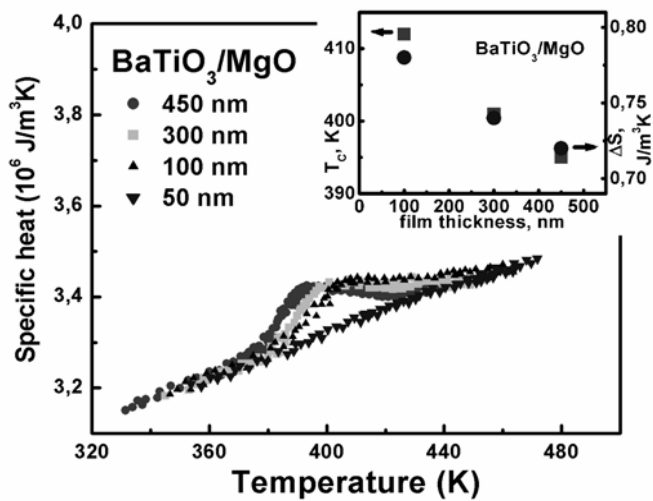

Fig. 5

Table 1. Estimation of the total residual strain in BaTiO$_3$/Mgo epitaxial films.

| h, nm | T$_C$ (K) | Δ T$_C$ (K), (size effect, poly-cryst. films) | Δ T$_C$ (K), (misfit strain, epitaxial films) | Thermal expansion strain | Total residual strain |
|---|---|---|---|---|---|
| **450** | **395** | **− 1** | **1** | **- 2,2·10$^{-4}$** | **2,8·10$^{-5}$** |
| **300** | **401** | **− 3** | **9** | | **2,5·10$^{-4}$** |
| **100** | **412** | **− 8** | **25** | | **7·10$^{-4}$** |
| **50** | **> 470** | **− 18** | **> 93** | | **> 2,6·10$^{-3}$** |